\documentclass{webofc}
\usepackage[varg]{txfonts}   
\usepackage{lineno}
\usepackage{xspace}
\begin{document}
\title{Constraining (anti)nuclei measurements relevant for astrophysics with ALICE}
%
%

\author{\firstname{Chiara} \lastname{Pinto}\inst{1}\fnsep\thanks{\email{chiara.pinto@cern.ch}} 
        \firstname{} \lastname{for the ALICE Collaboration}\inst{}
}

\institute{Technical University of Munich, TUM School of Natural Sciences, Physics Department, James-Franck-Stra{\ss}e 1, 85748 Garching b. M\"{u}nchen, Germany          }

\abstract{%
 Antinuclei can be produced in space either by collisions of high-energy cosmic rays with the interstellar medium or from the annihilation of dark matter particles stemming into standard model particles. High-energy hadronic collisions at accelerators create a suitable environment for producing light (anti)nuclei. Hence, studying the production of (anti)nuclei in pp collisions at the LHC can provide crucial insights into the production mechanisms of nuclear states in our Universe. Recent measurements of the production of (anti)nuclei in and out of jets, and as a function of rapidity in pp collisions at \mbox{$\sqrt{s}$ = 13 TeV} have been carried out with ALICE. The latter allow for the extrapolation of the nuclear production models at forward rapidity, region of interest for indirect searches of dark matter. 
Recent results on the annihilation cross-section of antinuclei are also discussed in the context of astrophysical measurements of cosmic ray flux. Such information is essential to study the different sources of antinuclei in our Universe and to interpret any future measurement of antinuclei in space.
}
\maketitle
\section{Introduction}
\label{intro}
Antinuclei in our Galaxy can be produced either by reactions between primary cosmic rays (CRs) and the interstellar medium (ISM), or by annihilation or decay of dark matter (DM) candidates. The relevant collisions, in the first case, are proton-proton and proton-nucleus, as both the CRs and the ISM mostly consist of protons (90\%) and alpha nuclei (8\%), and only for about 2\% of heavier nuclei~\cite{antideuteron}. 
According to the most accredited cosmological models, the expected flux of antinuclei measured in our Galaxy is given by the overlap of two contributions: (i) the flux due to antinuclei stemming from dark matter processes, and (ii) the flux due to antinuclei originating from cosmic ray interactions. The flux (i) is a factor $10^2-10^4$ higher at low kinetic energies ($E_{kin}\sim 10^{-1}$ GeV$/A$) than (ii)~\cite{He3absorption}, therefore the observation of an excess in the antinuclear flux at low $E_{kin}$ would represent a signal for dark matter annihilation or of more exotic physics beyond the standard model. To correctly interpret any future measurement of fluxes from antinuclei in our Universe, it is crucial to have precise knowledge of the production mechanism of antinuclei in the hadronic collisions relevant for astrophysics, and of the inelastic cross section of the annihilation processes that occur during the propagation of antinuclei in the Universe, before their detection. To this end, ALICE contributed performing measurements of production of (anti)nuclei in small systems (pp and p--Pb collisions), and measuring the inelastic cross section of antinuclei (namely $\overline{\rm d}$, $^3\overline{\rm H}$ and $^3\overline{\rm He}$) using the LHC as an antimatter factory and the ALICE detector as a target. 

\section{Production of nuclei}
\label{sec_production}
The production mechanism of light (anti)nuclei in high-energy hadronic collisions is a very debated topic in the scientific community. The experimental data are typically described using two phenomenological models: the statistical hadronization model (SHM) and baryon coalescence. In the former~\cite{SHM1}, light (anti)nuclei, as well as other hadron species, are assumed to be emitted by a source in local thermal and hadrochemical equilibrium with their abundances being fixed at chemical freeze-out, at a temperature of $T_{\mathrm{chem}}=156$ MeV~\cite{SHM5}. Multi-baryon states are assumed to be produced as compact multi-quark systems, which evolve towards the final configuration with typical timescales of 5 fm/$c$ or longer~\cite{SHM5}. This model provides an excellent description of the measured hadron yields in central nucleus--nucleus collisions~\cite{SHM5}. 
In the coalescence model~\cite{Coalescence1,Coalescence2}, multi-baryon states are assumed to be formed by coalescence of baryons that are close in phase space at kinetic freeze-out. In the simplest implementations of this model, the spatial correlations among nucleons are neglected and therefore the formation of a bound-state happens if the nucleons are close in momentum space, i.e., the difference between their momenta $\Delta p$ is below a given coalescence momentum $p_0$.
In the state-of-the-art implementations of the coalescence approach~\cite{Mahlein}, the quantum-mechanical properties of baryons and bound-states are taken into account and the coalescence probability is calculated from the overlap between the wave functions of individual (point-like) baryons and the Wigner density of the final-state cluster.

\subsection{Nuclei in jet and underlying event}
The measurement of the production of (anti)nuclei in jets and in the underlying event (UE) is a powerful tool to study the hadronization models, particularly coalescence. Indeed, by construction nucleons produced in jets have a tight constraint on the available phase-space, hence the coalescence probability in jets is expected to be enhanced with respect to the underlying event~\cite{dinjets}. 
\begin{figure}[!b]
\centering
\includegraphics[width=0.55\textwidth]{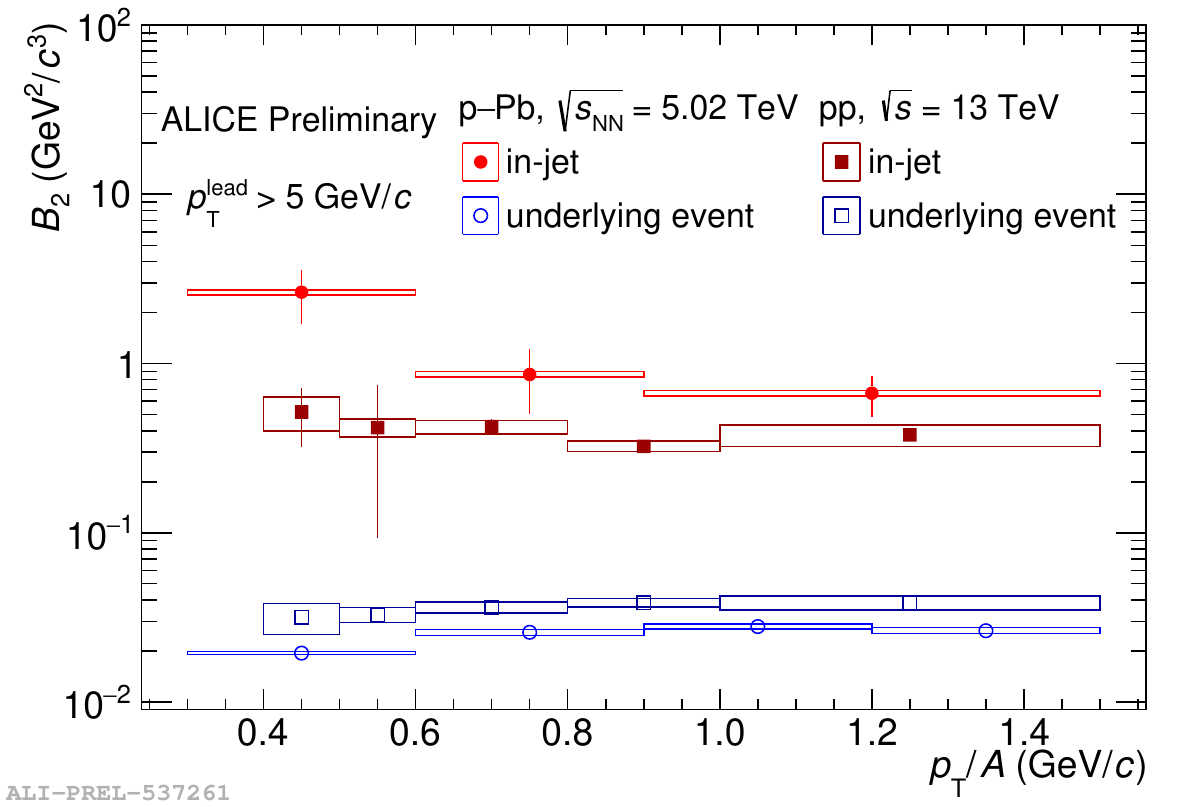}
\caption{Coalescence parameters $B_2$ in jets (red markers) and in the underlying event (blue markers), in pp~\cite{dinjets} (squares) and p--Pb (circles) collision systems, at $\sqrt{s}=$ 13 TeV and $\sqrt{s_{\rm {NN}}}=$ 5.02 TeV, respectively. }
\label{fig:B2_Jet_UE}       
\end{figure}
In Fig.~\ref{fig:B2_Jet_UE}, the coalescence parameters $B_2$ in jet and in the underlying event are shown as a function of $p_{\rm T}/A$, for pp~\cite{dinjets} and p--Pb collisions. Events with jets are selected by requiring that the leading track has a transverse momentum $p_{\rm T}^{\rm{lead}}>5$ GeV/$c$. An enhancement of the $B_2^{\rm{jet}}$ with respect to the $B_2^{\rm{UE}}$ of one order of magnitude is observed in both collision systems, being larger for p--Pb than for pp. In the underlying event, the coalescence probability is larger in pp collisions with respect to p--Pb ones, because the source size in the former is larger than in the latter system (\mbox{$\sim1$ fm} in pp collisions~\cite{ALICE:2018ysd} and $\sim 1.5$ fm in p--Pb collisions~\cite{ALICE:2019hdt}). The coalescence parameter in jets is larger in p--Pb than in the pp case, possibly because of the different particle composition of jets in the two collision systems. To confirm this hypothesis, though, more detailed studies of the identified particles produced in jets in different collision systems are needed.

\subsection{Rapidity dependence of the coalescence parameter}
\begin{figure}[!hbt]
\centering
\includegraphics[width=0.5\textwidth]{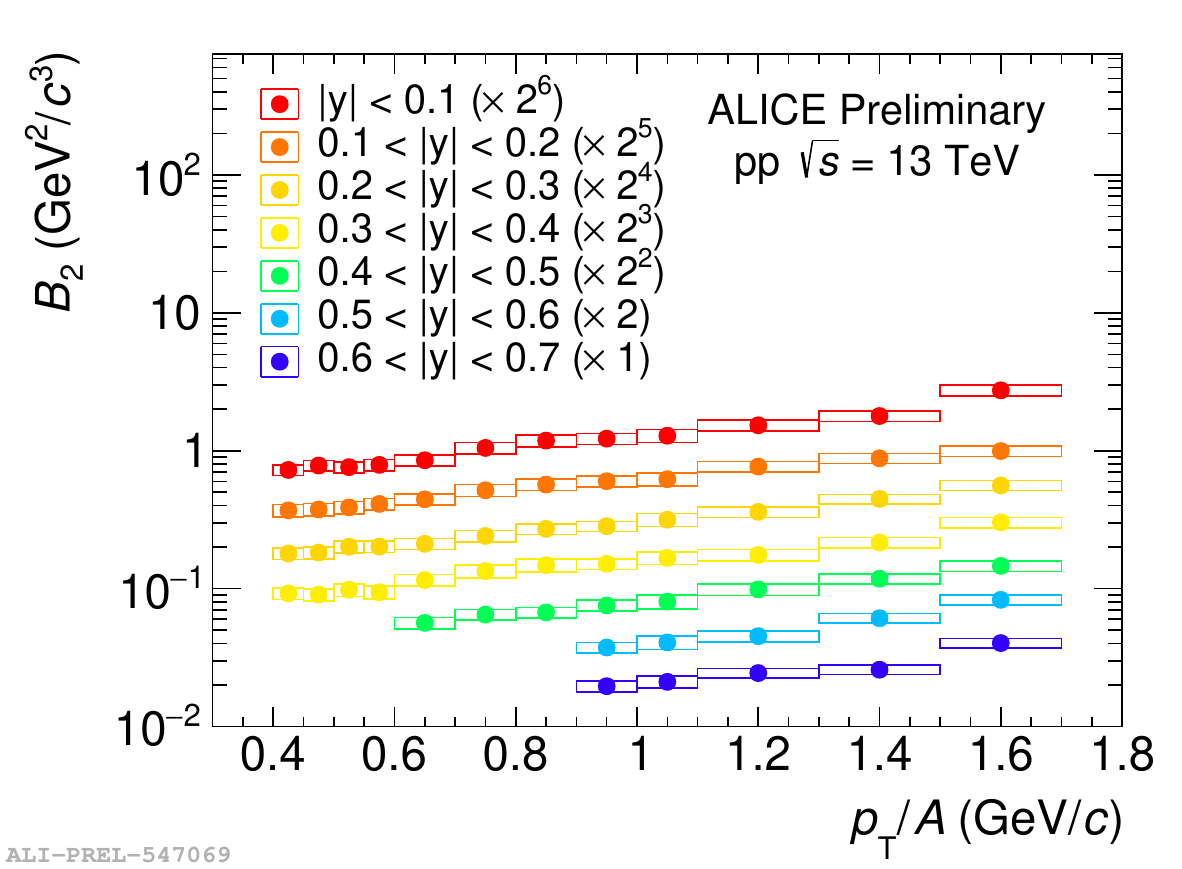}
\includegraphics[width=0.455\textwidth]{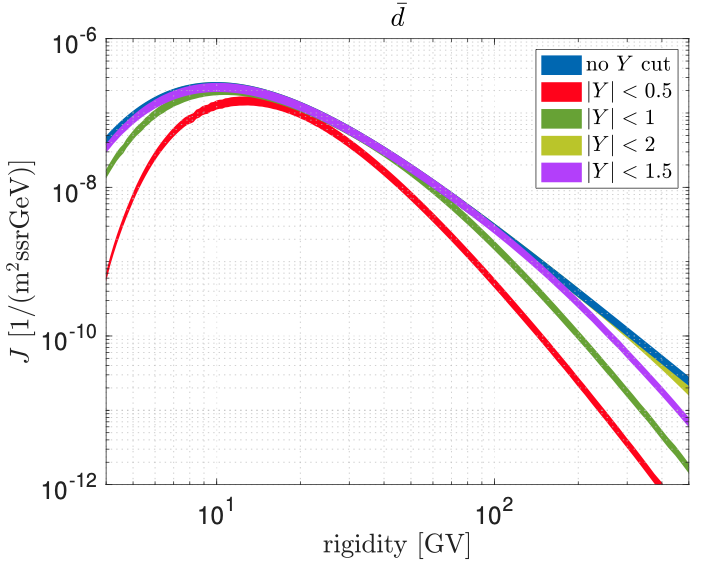}
\caption{\textit{Left.} Coalescence parameter $B_2$ measured in several rapidity intervals as a function of $p_{\rm T}/A$ in pp collisions at $\sqrt{s}=$ 13 TeV. \textit{Right.} Predicted flux of antideuterons from cosmic rays for different kinematical regions of rapidity~\cite{rapidity}. }
\label{fig:rapidity}      
\end{figure}

At the LHC, the production of (anti)nuclei is measured in the midrapidity region ($\left| y \right|<$ 0.5). 
These measurements are then utilised with various coalescence models to predict the flux of antinuclei from cosmic ray interactions at forward rapidity. However, the potential influence of a rapidity dependent production yield and coalescence probability of antinuclei has never been studied in details.
By measuring the production of antideuterons and antiprotons in several rapidity intervals in pp collisions at $\sqrt{s}=$13 TeV, up to $\left|y\right|=0.7$, the coalescence parameter $B_2$ (shown in the left panel of Fig.~\ref{fig:rapidity}) is obtained and then extrapolated at forward rapidity using multiple coalescence models. The predictions from these models are then given as input to the model of Ref.~\cite{Blum} to calculate the flux of antideuterons from cosmic rays~\cite{rapidity}, shown in the right panel of Fig.~\ref{fig:rapidity}.
This study reveals that the cosmic ray flux of antideuterons at low kinetic energies -- region of interest for astrophysics as expected to be dominated by the dark matter signal -- is dominated by nuclei produced in the rapidity region $\left| y \right|<$ 1.5, well in reach of the existing or future experimental facilities, such as LHCb~\cite{lhcb} and ALICE 3~\cite{alice3}. 

\section{Annihilation cross section}
\begin{figure}[h]
\centering
\includegraphics[width=0.55\textwidth]{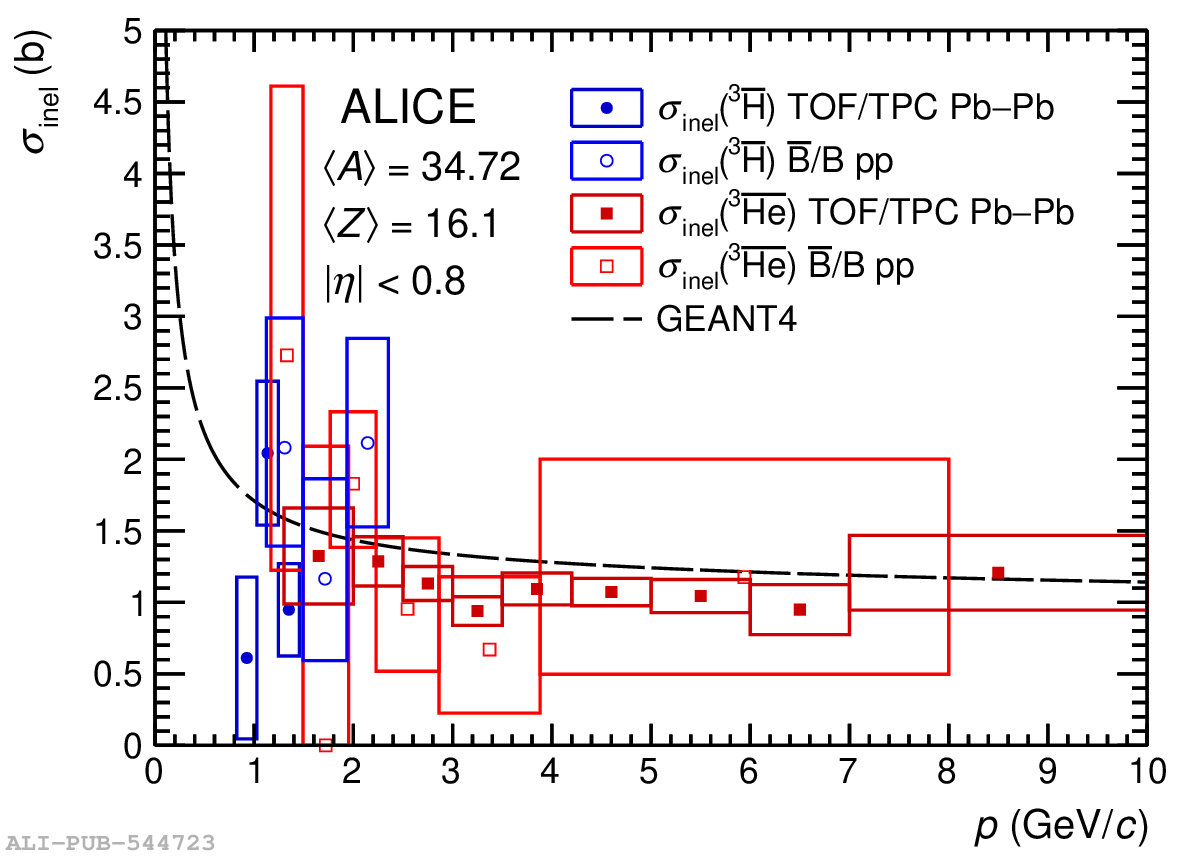}
\caption{Inelastic cross section for antitritons (blue markers) and (anti)$^3$He (red markers) on an average material element of the ALICE detector as a function of the momentum $p$ at which the interaction occurs~\cite{antitriton}. Dashed black lines represent the default GEANT4 parameterisations for antitritons. Boxes show the statistical and systematic uncertainties summed in quadrature.}
\label{fig:antit_sigma} 
\end{figure}

In their journey through the Galaxy, antinuclei may also annihilate before being detected. Hence, an important measurement is the inelastic cross section of antinuclei at the accelerator facilities. ALICE has measured the inelastic cross section of $\overline{\rm d}$~\cite{antideuteron}, $^3\overline{\rm He}$~\cite{He3absorption}, and recently also of $^3\overline{\rm H}$~\cite{antitriton}. 
The results for antinuclei with $A=$ 3 are shown in Fig.~\ref{fig:antit_sigma}, compared to the calculations based on the Glauber approach as implemented in the GEANT4 toolkit~\cite{Agostinelli:2002hh}, which is widely used in particle physics for the propagation of particles through the detector material. The comparison between the $\sigma_{\rm{inel}}(^3\overline{\rm H})$ and $\sigma_{\rm{inel}}(^3\overline{\rm He})$ results shows the feasibility of investigating the isospin dependence of inelastic interactions of antinuclei, specially with the larger data samples collected during the LHC Run 3. Moreover, the measurement of the $\sigma_{\rm{inel}}(^3\overline{\rm H})$ allows one to extend the measurement of the inelastic cross section of $^3\overline{\rm He}$ to a lower momentum region, which constitutes the most important region for the antinuclear annihilation for indirect dark-matter searches.

\end{document}